\documentclass{article}
\usepackage[T1]{fontenc}

\usepackage{spconf,amsmath,graphicx,hyperref}

\usepackage{amsmath,amssymb,amsfonts}
\usepackage{algorithmic}
\usepackage{graphicx}
\usepackage{textcomp}
\usepackage{stfloats}
\usepackage{url}
\usepackage{verbatim}
\usepackage{graphicx}
\usepackage{cite}
\usepackage{multirow}
\usepackage{lscape,array}
\usepackage{booktabs}
\usepackage{pifont}
\usepackage{ragged2e}
\usepackage{enumitem}
\usepackage{xcolor}
\usepackage[table]{xcolor}
\usepackage{hyperref}
\usepackage{balance}

\title{BUT Systems for Environmental Sound Deepfake Detection in the ESDD 2026 Challenge
}
\name{
Junyi Peng$^1$, Lin Zhang$^2$, Jin Li$^{1,3}$, Old\v{r}ich Plchot$^1$, Jan {\v{C}}ernock\'{y}$^1$
}

\address{
        $^1$Speech@FIT, Brno University of Technology, Czechia
        $^2$Johns Hopkins University, USA \\
        $^3$Department of EEE, The Hong Kong Polytechnic University, Hong Kong SAR
    }

\begin{document}
% \ninept
\maketitle

\begin{abstract}
This paper describes the BUT submission to the ESDD 2026 Challenge, specifically focusing on Track 1: Environmental Sound Deepfake Detection with Unseen Generators. To address the critical challenge of generalizing to audio generated by unseen synthesis algorithms, we propose a robust ensemble framework leveraging diverse Self-Supervised Learning (SSL) models. We conduct a comprehensive analysis of general audio SSL models (including BEATs, EAT, and Dasheng) and speech-specific SSLs. These front-ends are coupled with a lightweight Multi-Head Factorized Attention (MHFA) back-end to capture discriminative representations. Furthermore, we introduce a feature domain augmentation strategy based on distribution uncertainty modeling to enhance model robustness against unseen spectral distortions. All models are trained exclusively on the official EnvSDD data, without using any external resources. Experimental results demonstrate the effectiveness of our approach: our best single system achieved Equal Error Rates (EER) of 0.00\%, 4.60\%, and 4.80\% on the Development, Progress (Track 1), and Final Evaluation sets, respectively. The fusion system further improved generalization, yielding EERs of 0.00\%, 3.52\%, and 4.38\% across the same partitions.
\end{abstract}
\section{Introduction}
Unlike speech deepfake detection (e.g., ASVspoof~\cite{wang24_asvspoof}), which has been extensively studied, Environmental Sound Deepfake Detection (ESDD) remains an underexplored area~\cite{yin2025envsdd}. Environmental sounds are inherently unstructured and exhibit significantly higher diversity in spectro-temporal patterns compared to human speech. This complexity is further exacerbated by domain shifts, as detection models trained on specific generators often fail to generalize to unseen ones—different vocoders and diffusion processes leave distinct forensic artifacts.

The ESDD 2026 Challenge~\cite{yin2025esdd} addresses this critical gap by evaluating systems under strict generalization protocols (Track 1) and low-resource constraints (Track 2). The primary challenge lies in learning robust discriminative features that persist across unknown synthesis algorithms without relying on generator-specific bias.

In this paper, we present the \textbf{BUT} submission to ESDD 2026 Track1. Instead of relying on massive external datasets, our approach focuses on maximizing the representation power of Self-Supervised Learning (SSL). We hypothesize that large-scale pre-trained audio models encode rich acoustic information that can be effectively repurposed for artifact detection. We conduct a systematic study of general audio SSLs (BEATs~\cite{chen2022beats}, EAT~\cite{chen2024eat}, Dasheng~\cite{dinkel2024scaling}) versus speech-specific SSLs (e.g. WavLM~\cite{chen2022wavlm}). To prevent overfitting to the training generators, we introduce a \textbf{Multi-Head Factorized Attention (MHFA)} back-end~\cite{peng2022attention} along with a novel feature domain augmentation~\cite{li2022uncertainty} strategy.

Our main contributions are summarized as follows:
\begin{itemize}[noitemsep]
    \item We conduct a systematic comparison between speech-specific and general audio SSL models for environmental sound deepfake detection, showing that the latter are consistently superior.
    \item We adapt the MHFA back-end to the env-deepfake setting with hierarchical layer-wise aggregation.
    \item We integrate DSU-based feature-domain augmentation into MHFA and show improved robustness under strict Track 1 protocols.
\end{itemize}

\section{Proposed Method}
\label{sec:method}

\subsection{Data and Framework}
The proposed system is implemented within the WeDefense~\cite{zhang2026wedefense}\footnote{\url{https://github.com/zlin0/wedefense}} framework, an open-source toolkit tailored for robust deepfake detection. We utilize the official EnvSDD dataset as the exclusive source for both training and development. To ensure consistency across different back-ends, we adopt a unified preprocessing protocol: all training clips are randomly cropped to 4 seconds, matching the fixed duration used during testing.

\subsection{Hierarchical SSL Feature Extraction}
Self-Supervised Learning models, such as WavLM, HuBERT, and BEATs, encode rich acoustic information across their layers. Lower layers typically capture raw spectral details, while upper layers encode more semantic or structural information~\cite{10887574}. For the task of deepfake detection, relying solely on the last layer is often sub-optimal, as forensic artifacts introduced by neural vocoders (e.g., phase discontinuities or metallic buzzing) are often preserved in intermediate representations.

Therefore, instead of using only the final hidden state, we utilize a weighted sum of all $L$ transformer layers. Let $X \in \mathbb{R}^{L \times T \times D}$ be the output features from all layers of the SSL encoder, where $T$ is the number of frames and $D$ is the feature dimension. We learn layer-specific weights to aggregate these features dynamically, allowing the back-end to focus on the most discriminative information.

\begin{figure*}
    \centering
    \includegraphics[width=0.9\linewidth]{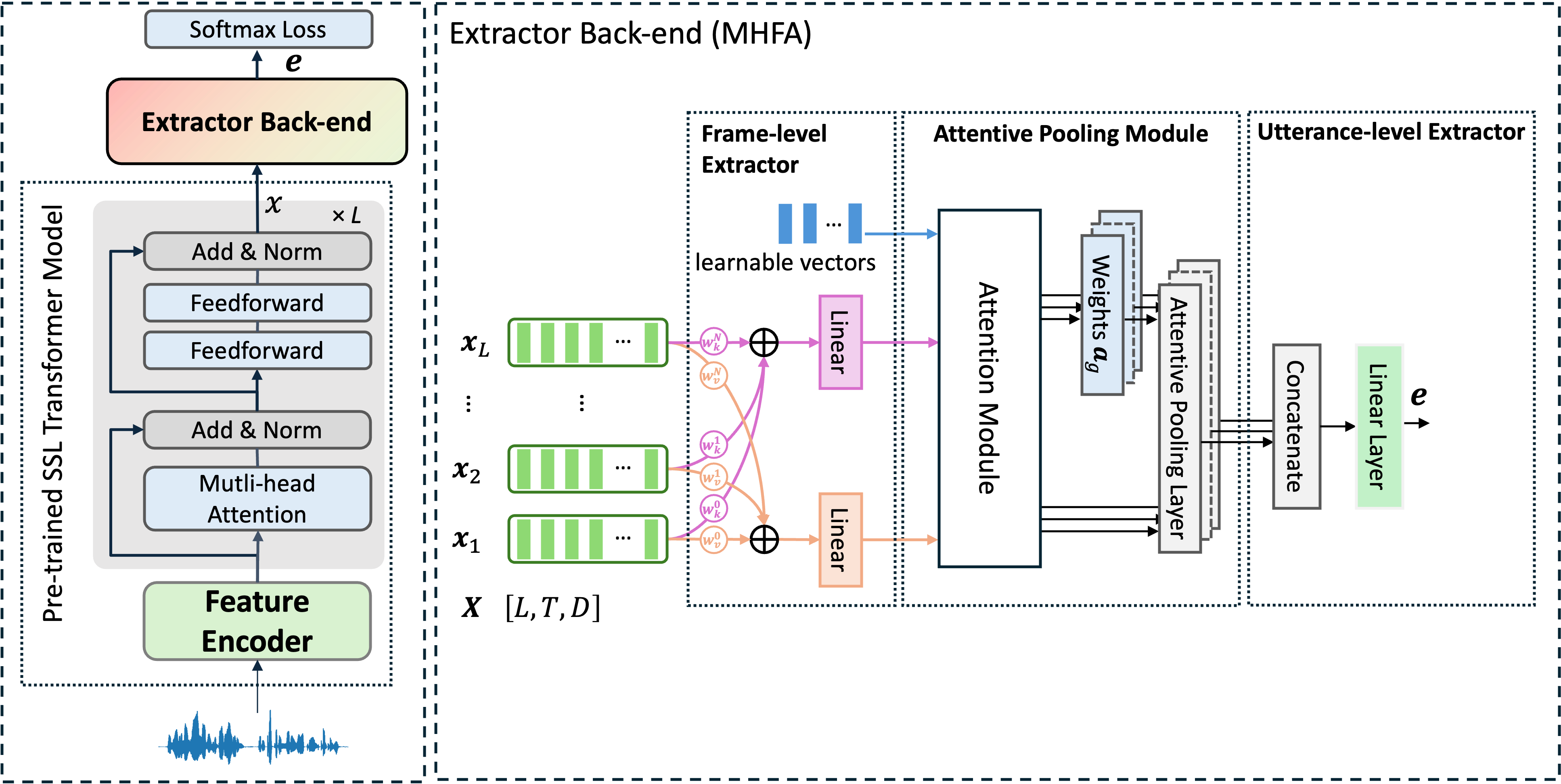}
    \caption{Overview of the proposed Environmental Sound Deepfake Detection framework. The left part depicts the pre-trained SSL backbone, contrasting Audio SSLs (using Patch Embedding and Linear Projection) with Speech SSLs (using a 7-layer CNN Encoder), utilizing a weighted sum of all transformer layers for hierarchical feature extraction. The right part details the Multi-Head Factorized Attention (MHFA) back-end, where Feature Domain Augmentation (DSU) is injected into the Value stream to enhance robustness against unseen generators.}
    \label{fig:enter-label}
\end{figure*}

\subsection{Multi-Head Factorized Attention (MHFA)}
To effectively aggregate the temporal frame-level features into a global utterance-level embedding, we employ the \textbf{Multi-Head Factorized Attention (MHFA)} mechanism. Unlike standard attention, which uses a single linear projection, MHFA factorizes the aggregation process into two independent streams: a \textit{Key} stream ($K$) and a \textit{Value} stream ($V$), as shown in Fig~\ref{fig:enter-label}.

Specifically, we define two separate sets of learnable layer weights, $w^k \in \mathbb{R}^L$ and $w^v \in \mathbb{R}^L$. These weights are normalized via softmax to compute the weighted sum of the SSL layer outputs $Z_l$:
\begin{equation}
    K_{feat} = \sum_{l=1}^{L} \text{softmax}(w^k_l) Z_l, \quad V_{feat} = \sum_{l=1}^{L} \text{softmax}(w^v_l) Z_l
\end{equation}
These aggregated features are then projected into a lower dimension $D_{cmp}$ using linear layers $W_k$ and $W_v$:
\begin{equation}
    K = K_{feat} W_k, \quad V = V_{feat} W_v
\end{equation}
The attention weights $A$ are computed from the Query stream, while the content to be aggregated comes from the Value stream. This factorization allows the model to learn \textit{where} to look using $K$ independently of \textit{what} to extract using $V$. For $H$ heads, the output is pooled as:
\begin{equation}
    A = \text{softmax}(K W_{att}, \text{dim}=1)
\end{equation}
\begin{equation}
    Embedding = \text{Pooling}(V \odot A)
\end{equation}
Finally, a fully connected layer maps the concatenated head outputs to the final embedding $e$. This embedding is then processed by a standard binary classification head distinguishing between \textit{bonafide} and \textit{spoof} audio.

\subsection{MHFA with DSU (Feature Domain Augmentation)}
To tackle the challenge of unseen generators (Track 1), we integrate a Feature Domain Augmentation strategy directly into the MHFA back-end~\cite{li2022uncertainty}, termed \textbf{MHFA-DSU}. This method is based on the concept of Distribution Uncertainty (DSU), which hypothesizes that domain shifts can be simulated by perturbing the feature statistics (mean and variance) of the training data.

We apply DSU specifically to the Value stream ($V_{feat}$) before the linear projection. During training, with a probability $p$, we model the feature statistics as distributions rather than deterministic values. For an input feature map $x$ (corresponding to $V_{feat}$), we compute the instance-level mean and standard deviation across the temporal dimension \cite{li2022uncertainty}. This operation essentially ``jitters'' the global style and channel characteristics of the audio representation while preserving the local content, forcing the network to learn features that are invariant to global statistical shifts caused by different vocoders.

\section{Experiments}

\subsection{Experimental Setup}
Our models were implemented using PyTorch and trained on AMD Instinct MI200 GPUs. The training process was configured with a maximum of 8 epochs. We utilized the AdamW optimizer with $\beta_1=0.9$, $\beta_2=0.999$, and a weight decay of $1.0 \times 10^{-4}$. The batch size was set to 128.

We employed a differential learning rate strategy to fine-tune the SSL front-end and the MHFA back-end effectively. The base learning rate was initialized at $5.0 \times 10^{-4}$ and decayed to a final learning rate of $1.0 \times 10^{-5}$ using a Cosine Annealing scheduler. To prevent catastrophic forgetting of the pre-trained representations, the learning rate for the SSL front-end was scaled by a factor of 0.05 relative to the base learning rate. A warmup period of 2 epochs was applied at the beginning of training.

For the MHFA back-end configuration, we set the number of attention heads (\texttt{head\_nb}) to 32, the embedding dimension (\texttt{embed\_dim}) to 256, and the compression dimension (\texttt{compression\_dim}) to 128.

\subsection{Results}
We report the performance using EERs. Table \ref{tab:results} compares our proposed systems against the official baselines.

\begin{table}[ht]
\centering
\caption{Performance comparison on ESDD 2026. $*$ denotes the system uses MHFA-DSU}
\label{tab:results}
\begin{tabular}{clccc}
\toprule
\textbf{ID} & \textbf{System} & \textbf{Dev} & \textbf{Prog} & \textbf{Eval} \\
\midrule
1  & Baseline (AASIST)~\cite{yin2025envsdd}          & 0.92 & 15.26 & 15.02 \\
2  & Baseline (BEATs+AASIST)~\cite{yin2025envsdd}   & 0.10 & 14.21 & 13.20 \\
\midrule
3  & WavLM Base +                                   & 4.75 &  -    &  -    \\
4  & Dasheng-0.6B                                   & 0.27 &  -    &  -    \\
5  & Dasheng-1.2B                                   & 0.33 &  -    &  -    \\
6  & BEATs                                          & 0.00 & 7.10 &  -    \\
7  & BEATs (FT-AS2M-CPT1)                           & 0.00 & 6.06 &  -    \\
8  & BEATs (FT-AS2M-CPT2)                           & 0.00 & 5.65 &  -    \\
9  & MAE++ Base                                     & 0.00 & 6.46 &  -    \\
10 & EAT Base                                       & 0.00 & 5.41 &  -    \\
11 & EAT Base\_FT(AS2M)                             & 0.00 & 4.60 & 4.80 \\
12 & EAT Base\_FT(AS2M)*                            & 0.00 & 4.77 & 4.80 \\
13 & EAT Large                                      & 0.00 & 5.68 &  -    \\
14 & EAT Large\_FT(AS2M)                            & 0.00 & 4.60 & 5.50 \\
\midrule
- & Fusion (11+12)  & 0.00 & 3.99 & -\\
- & Fusion (11+12+14)                                        & 0.00 & 3.52 & 4.38 \\
\bottomrule
\end{tabular}
\end{table}

\subsection{Analysis}
The experimental results, summarized in Table \ref{tab:results}, demonstrate the effectiveness of our proposed approach.

\textbf{Comparison with Baselines:} All our submitted systems significantly outperform the official baselines. Our best single system (EAT Base\_FT) achieves an EER of 4.80\% on the Evaluation set, representing a relative improvement of over 63\% compared to the BEATs+AASIST baseline (13.20\%). This validates the superiority of our hierarchical feature extraction and MHFA back-end.

\textbf{General Audio vs. Speech SSLs:} We observe a distinct performance gap between speech-specific SSLs (WavLM) and general audio SSLs (BEATs, EAT, Dasheng). WavLM yields a high EER of 4.75\% on the development set, whereas general audio models consistently achieve near-zero errors. This confirms that general audio models, pretrained on diverse sound events, possess stronger representation capabilities for unstructured environmental sounds compared to speech-tailored models.

\textbf{Impact of Pre-training Data:} Further fine-tuning or continued pre-training on the AudioSet-2M (AS2M) dataset~\cite{gemmeke2017audio} proves beneficial. For instance, fine-tuning EAT Base on AS2M reduces the Progress EER from 5.41\% to 4.60\%. Similarly, continued pre-training of BEATs (CPT2) improves performance from 7.10\% to 5.65\%. This suggests that exposure to diverse real-world acoustic environments during training enhances the model's ability to distinguish authentic sounds from synthetic artifacts.

\textbf{Effectiveness of MHFA-DSU:} The integration of Distribution Uncertainty (DSU) based augmentation (denoted by *) maintains competitive performance (4.80\% Eval EER) while theoretically enhancing robustness against domain shifts. By simulating unseen feature statistics, MHFA-DSU helps the model generalize better to the unseen generators in Track 1.

\textbf{Fusion Strategy:} Finally, our ensemble system achieves the best overall performance with an EER of 3.52\% on the Progress set and 4.38\% on the Evaluation set. This fusion combines the scores of three robust systems: EAT Base\_FT(AS2M), EAT Base\_FT(AS2M) with MHFA-DSU, and EAT Large\_FT(AS2M).

\section{Conclusion}
In this work, we presented a robust detection system for environmental sound deepfakes. By leveraging an ensemble of SSL models (BEATs, EAT, Dasheng) and a lightweight MHFA back-end, we significantly improved generalization to unseen generators.

% \section{Acknowledgment}
% The work was supported by European Defense Fund's project No. 101168083 "ARCHER", Czech Ministry of Interior project No. VJ01010108 "ROZKAZ", Ministry of Education, Youth and Sports of the Czech Republic (MoE) through the OP JAK project "Linguistics, Artificial Intelligence and Language and Speech Technologies: from Research to Applications"(ID: CZ.02.01.01/00/23\_020/0008518) and Horizon 2020 Marie Sklodowska-Curie grant ESPERANTO, No. 101007666. Computing on IT4I supercomputer was supported by the Czech Ministry of Education, Youth and Sports through the e-INFRA CZ (ID: 90254).

% \balance
{\footnotesize
\bibliographystyle{IEEEtran}
\bibliography{mybib}
}
\end{document}